\documentclass[%
 reprint,
 onecolumn,
 amsmath,amssymb,
 aps,
prab,
]{revtex4-2}

\usepackage{graphicx}
\usepackage{float}
\usepackage[caption = false]{subfig}
\usepackage{dcolumn}
\usepackage{bm}

\begin{document}

\preprint{APS/123-QED}

\title{Final multiplicity of a QED cascade in generalized Heitler  model}

\author{Y.~V. Selivanov}
\author{A.~M. Fedotov}
\affiliation{National Research Nuclear University MEPhI, Moscow, 115409, Russia}


\begin{abstract}
We consider a generalized Heitler model for QED cascade. An exact formula for the final number of leptons is obtained by solving the kinetic equations. We demonstrate that in such a model the final number of leptons does not depend on photon and lepton free paths. We derive approximate formulas for the main characteristics of cascades at high energy, including the final number of leptons and the cascade depth. We show that in general the final number of leptons is asymptotically proportional to the energy of seed particle. It is also demonstrated how the original Heitler model is reproduced as a special case.
\end{abstract}

\maketitle


\section{Introduction}
\label{sec1}
QED cascade (also called electromagnetic cascade) is a chain of successive events of hard photon emission and electron-positron pair photoproduction.
This happens when a high-energy photon or lepton enters media or strong external field producing a bunch of secondary particles. Cascades have been widely studied as a part of cosmic air showers \cite{auger1939extensive,gaisser1990cosmic} and as a strong-field QED phenomenon \cite{akhiezer1994kinetic,anguelov1999electromagnetic,
fedotov2010limitations,elkina2011qed,
bulanov2013electromagnetic}. In the latter case it was proposed to distinguish between ordinary and self-sustained cascades, also called showers and avalanches, respectively \cite{mironov2014collapse,narozhny2015quantum,fedotov2023advances}. The main difference is the energy source. For self-sustained cascades (avalanches) energy comes from the field, while for ordinary ones (showers) it eventually comes from the seed particle. Therefore in the latter case the dependence of cascade multiplicity on the seed particle energy is of fundamental interest. In this case as cascade multiplicity grows, the energy per particle decreases reaching a certain threshold - critical energy $E_0$ when photoproduction  stops. This limits the number of leptons produced in such cascades.
 
A cascade theory of showers was developed independently by Carlson and Oppenheimer \cite{CarlsonOppenheimer} and Bhabha and Heitler \cite{BhabhaHeitler} and was further advanced by many authors (see, e.g., \cite{LandauRumer,heitler1948quantum,gaisser1990cosmic}). Obtaining exact analytical solutions in cascade theory is challenging and for fully realistic models is hardly possible. However, useful qualitative results can be obtained through simple models. The simplest toy model for QED cascade is the Heitler model  \cite{heitler1948quantum}, in which all the decays (photon emissions and pair photoproductions) happen after passing the same free path~\footnote{In this paper we define free paths up to the factor $\ln{2}$.} $L$ and with equal energy splitting between the secondary particles.

In this model at high seed particle energy $\varepsilon_0$ the final number of leptons can be approximated as
\begin{equation}\label{Heitler_Ne}
    N_e \approx \frac{2}{3} \, \cfrac{\varepsilon_0}{E_0},
\end{equation}
and is achieved at depth
\begin{equation}\label{Heitler_tm}
    t_m \approx L \ln\left(\frac{\varepsilon_0}{E_0}\right).
\end{equation}

\begin{figure}[t]
    \centering
\includegraphics[width=0.8\textwidth]{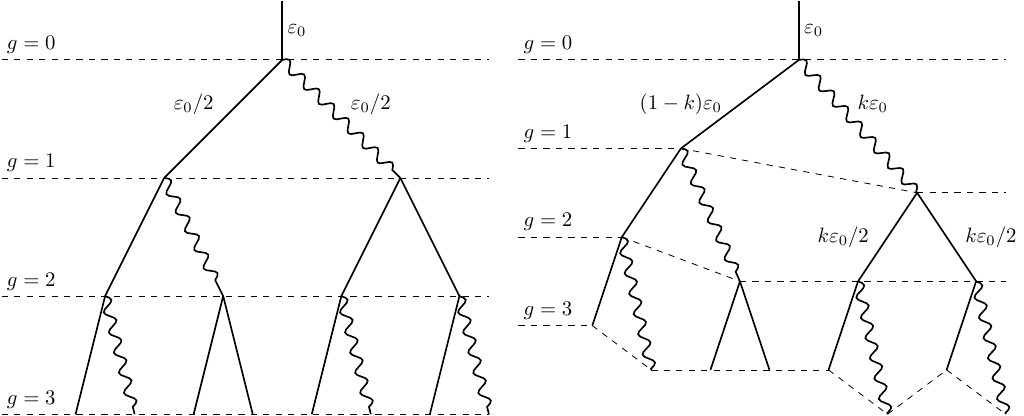}
    \caption{Schematics of the electron-seeded QED cascade in  original (left) and generalized (right) Heitler model. Straight and wavy lines correspond to leptons and photons, respectively. Their different heights represent different free paths, $g$ is the number of generation.}
    \label{fig:schematics}
\end{figure}

In this paper we consider a generalization of the Heitler model. Namely, we introduce a new parameter $k$, energy transfer coefficient for photon emission, so that a lepton with energy $\varepsilon$ emits photon with energy $k\varepsilon$ (in Heitler model $k = 1/2$). In addition, we allow for different  free paths $L_e, L_{\gamma}$ for leptons and photons, respectively. Differences between the original Heitler model and our generalization are illustrated in Fig.~\ref{fig:schematics}. Note that since we have in mind mostly QED cascades, our generalization differs from those suggested for the hadronic component of extensive air showers \cite{matthews2005heitler,
mariazzi2018phenomenology}. Using the kinetic equations method from cascade theory we obtain energy distributions for photons and leptons and use them to derive analytically the final number of leptons $N_e$ and the cascade depth $t_m$.

\section{Cascade equations}
\label{sec2}
Let us start with a  brief review of kinetic equations in cascade theory, specifically for QED cascades. Assuming the number of particles is large, we consider lepton and photon energy distributions
\begin{eqnarray}
    f_e(\varepsilon, t) = \cfrac{dN_e(\varepsilon, t)}{d\varepsilon}, ~
    f_{\gamma}(\varepsilon, t) = \cfrac{dN_{\gamma}(\varepsilon, t)}{d\varepsilon},
\end{eqnarray}
where $N_e(\varepsilon,t)$ and $N_{\gamma}(\varepsilon,t)$ are the numbers of leptons and photons with energies up to $\varepsilon$ at depth $t$, respectively.

The QED processes are described by the differential rates $W_{e \rightarrow \gamma}(\varepsilon, \varepsilon')$ and $W_{\gamma \rightarrow e}(\varepsilon, \varepsilon')$ giving the probabilities for photon emission and pair photoproduction per unit depth and unit energy range of the final particle, respectively. In case of $W_{e \rightarrow \gamma}(\varepsilon, \varepsilon')$, $\varepsilon$ and $\varepsilon'$ denote the initial lepton energy and the energy of emitted photon, respectively. And in case of $W_{\gamma \rightarrow e}(\varepsilon, \varepsilon')$, $\varepsilon$ and $\varepsilon'$ are the energies of the initial photon and of the produced electron, respectively. 

With this notation the kinetic equations for QED cascade take the form (see, e.g.,  \cite{LandauRumer,akhiezer1994kinetic,elkina2011qed})
\begin{equation}
\label{eq:casc_eq}
\begin{aligned}
        \cfrac{\partial f_{e}(\varepsilon, t)}{\partial t} = &
        \int_{\varepsilon}^{\infty}f_{e}(\varepsilon', t) \, W_{e \rightarrow \gamma}(\varepsilon',\varepsilon' - \varepsilon) d\varepsilon' 
        + 2 \, \int_{\varepsilon}^{\infty}f_\gamma(\varepsilon', t)\, W_{\gamma \rightarrow e}(\varepsilon', \varepsilon) d\varepsilon'
        - \int_{0}^{\varepsilon}f_{e}(\varepsilon, t)\, W_{e \rightarrow \gamma}(\varepsilon, \varepsilon') d\varepsilon', \\
        \cfrac{\partial f_\gamma(\varepsilon, t)}{\partial t} = & \int_{\varepsilon}^{\infty}f_{e}(\varepsilon', t) \, W_{e \rightarrow \gamma}(\varepsilon', \varepsilon) d\varepsilon' - \int_{0}^{\varepsilon}f_\gamma(\varepsilon, t)\, W_{\gamma \rightarrow e}(\varepsilon, \varepsilon') d\varepsilon'.
\end{aligned}
\end{equation}
The initial conditions depend on a scenario, for example for a cascade seeded by single lepton with energy $\varepsilon_0$ they read
\begin{equation}\label{eq:ini_conds}
 f_e(\varepsilon, 0) = \delta(\varepsilon-\varepsilon_0), ~f_{\gamma}(\varepsilon, 0) = 0.
\end{equation}

Equations \eqref{eq:casc_eq} along with \eqref{eq:ini_conds} completely determine the cascade dynamics, in particular the final number of leptons and the cascade depth in terms of the initial energy $\varepsilon_0$.

\section{Generalized Heitler model}
\label{sec3}

The explicit forms of the rates $W_{e \rightarrow \gamma}(\varepsilon, \varepsilon')$ and $W_{\gamma \rightarrow e}(\varepsilon, \varepsilon')$ are derived from cross sections and are specific to particular scenario. For example, in a medium $W_{e \rightarrow \gamma}(\varepsilon, \varepsilon')$ is defined by the cross section of Bremsstrahlung and $W_{\gamma \rightarrow e}(\varepsilon, \varepsilon')$ is defined by the cross section of Bethe-Heitler process \cite{berestetskii1982quantum}. The above rates are also known for QED processes in constant and plane wave electromagnetic fields \cite{ritus1985quantum}.

However, for the sake of simplicity and generality in this paper we adopt a toy model by setting
\begin{eqnarray} \label{W_eg}
    W_{e \rightarrow \gamma}(\varepsilon, \varepsilon') = \cfrac{1}{L_e} \, \delta(\varepsilon'-k\varepsilon ), ~
    W_{\gamma \rightarrow e}(\varepsilon, \varepsilon') =  \cfrac{1}{L_{\gamma}} \, \delta(\varepsilon'-\varepsilon / 2).
\end{eqnarray}
This generalizes the Heitler model first by making energy transfer to the emitted photon arbitrary and second by allowing for different free paths for leptons and photons. Unlike for photon emission, within the deterministic model charge symmetry requires equal energy splitting between the photoproduced electron and positron. Note that in writing $W_{\gamma \rightarrow e}$ in Eq.~\eqref{W_eg} we assume $\varepsilon\geqslant E_0$, where $E_0$ is the pair photoproduction threshold.

By substituting Eqs.~\eqref{W_eg} into Eqs.~\eqref{eq:casc_eq} we arrive at
\begin{equation}
\label{kinetic_equations}
\begin{aligned}
    \cfrac{\partial f_{e}(\varepsilon,t)}{\partial t} = & \cfrac{1}{L_e \, (1-k)}\, f_{e}\left(\frac{\varepsilon}{1 - k},t\right) + \cfrac{4}{L_{\gamma}}\, f_{\gamma}(2 \, \varepsilon,t) - \cfrac{1}{L_e}\, f_{e}(\varepsilon,t),\\
\cfrac{\partial f_{\gamma}(\varepsilon,t)}{\partial t} = &\cfrac{1}{L_e \, k}\, f_{e}\left(\frac{\varepsilon}{k},t\right) - \cfrac{1}{L_{\gamma}}\, f_{\gamma}(\varepsilon,t). 
\end{aligned}
\end{equation}
We emphasize that Eqs.~\eqref{kinetic_equations} are only valid for energies $\varepsilon\geqslant E_0$. In what follows we adopt the initial conditions \eqref{eq:ini_conds}.

\section{Distribution functions at \bm{$\varepsilon \geqslant E_0$}}
\label{sec4}
Equations \eqref{kinetic_equations} are first order linear differential with respect to depth $t$ and functional with respect to energy $\varepsilon$. In order to solve them we apply the Mellin transform $\mathcal{M}$ over $\varepsilon$ using the following definition
\begin{equation}\label{Mellin transform}
    \hat{f}(s) = \mathcal{M}[f](s) = \int_0^{\infty}\varepsilon^sf(\varepsilon)d\varepsilon.
\end{equation}
The inverse transform is then given by
\begin{eqnarray}\label{Mellin transform inv}
    f(\varepsilon) = \cfrac{1}{2\pi i}\int_{\sigma - i\infty}^{\sigma + i\infty}\cfrac{1}{\varepsilon^{s+1}}\hat{f}(s)ds.
\end{eqnarray}
Integral in Eq.~\eqref{Mellin transform inv} is taken along a vertical line in complex plane and $\sigma$ is such that $\hat{f}(s)$ is analytical in the halfplane $\Re(s) > \sigma$.

By applying the Mellin transform to Eqs.~\eqref{kinetic_equations} and \eqref{eq:ini_conds} we arrive at
\begin{align}
    \cfrac{\partial \hat{f_{e}}(s, t)}{\partial t} = &\cfrac{(1-k)^s - 1}{L_e}\, \hat{f_{e}}(s, t) + \cfrac{(1/2)^{s-1}}{L_{\gamma}}\, \hat{f_{\gamma}}(s, t),\\
    \cfrac{\partial \hat{f_{\gamma}}(s, t)}{\partial t} = &\cfrac{k^s}{L_e}\, \hat{f_e}(s, t) - \cfrac{1}{L_{\gamma}}\, \hat{f_{\gamma}}(s, t),\\
    \hat{f_e}(s, 0) = &\varepsilon_0^s, ~ \hat{f_{\gamma}}(s, 0) = 0.
\end{align}
These are the first order linear differential equations with parameter $s$ and as such can be solved analytically. Solution for the functions $\hat{f}_e$ and $\hat{f}_{\gamma}$ comes in form of linear combinations of the exponentials:
\begin{align}\label{eq:f_e}
    \hat{f_{e}}(s, t) =& \cfrac{\varepsilon_0^s}{\lambda_1 - \lambda_2} \, \left[ \left(\cfrac{1}{L_{\gamma}} + \lambda_1 \right) \, e^{\lambda_1 t} - \left(\cfrac{1}{L_{\gamma}} + \lambda_2 \right) \, e^{\lambda_2 t} \right], \\
    \label{eq:f_gamma}
    \hat{f_{\gamma}}(s, t) =& \cfrac{\varepsilon_0^s \, k^s}{\lambda_1 - \lambda_2} \, \cfrac{1}{L_e}\left(e^{\lambda_1 t} - e^{\lambda_2 t} \right),\\
    \lambda_{1, 2} =& ~\cfrac{1}{2}\left( \cfrac{(1-k)^s-1}{L_e} - \cfrac{1}{L_{\gamma}}\right) \pm \cfrac{1}{2}\sqrt{\left( \cfrac{(1-k)^s-1}{L_e} + \cfrac{1}{L_{\gamma}}\right)^2 + \cfrac{8}{L_e L_{\gamma}}\,\left(\cfrac{k}{2}\right)^s}.
\end{align}

Expressions \eqref{eq:f_e} and \eqref{eq:f_gamma} can be expanded in whole powers of $(1-k)^s$ and $(k/2)^s$:
\begin{eqnarray} 
    \hat{f}_e(s,t) = \varepsilon_0^s \, \sum_{p, l = 0}^{\infty}A_{pl}\left(t, L_e, L_{\gamma}\right) (1-k)^{ps} (k/2)^{ls}, \label{f_hat_e_expansion}\\
    \hat{f}_{\gamma}(s,t) = \varepsilon_0^s \, k^s \, \sum_{p, l = 0}^{\infty}B_{pl}\left(t, L_e, L_{\gamma}\right) (1-k)^{ps} (k/2)^{ls}.\label{f_hat_g_expansion}
\end{eqnarray}
The explicit form of the coefficients $A_{pl}$ and $B_{pl}$ is obtained in  Appendix~\ref{sec:app}, see Eqs.~\eqref{Apl} and \eqref{Bpl}, respectively. 

From the expansions \eqref{f_hat_e_expansion} and \eqref{f_hat_g_expansion} it follows that functions $\hat{f}_e$ and $\hat{f}_{\gamma}$ are analytical in complex variable $s$. This means that in integral~\eqref{Mellin transform inv} $\sigma$ can be set to zero:
\begin{align}\label{f_e general}
        \nonumber f_e(\varepsilon, t) =&  \cfrac{1}{2\pi i}\int_{-i\infty}^{+i\infty} \cfrac{1}{\varepsilon^{s+1}}\, \hat{f_{e}}(s, t)ds = \sum_{p,l=0}^{\infty}A_{pl}\left(t, L_e, L_{\gamma}\right) \, \cfrac{1}{2\pi\varepsilon}\int_{-\infty}^{+\infty}\exp\left[i\omega\ln\left(\cfrac{\varepsilon_0(1-k)^p(k/2)^l}{\varepsilon}\right)\right]d\omega = \\
        =& \sum_{p,l=0}^{\infty}A_{pl}\left(t, L_e, L_{\gamma}\right) \, \cfrac{1}{\varepsilon} \, \delta\left( \cfrac{\varepsilon_0(1-k)^p(k/2)^l}{\varepsilon} \right) = \sum_{p,l=0}^{\infty}A_{pl}\left(t, L_e, L_{\gamma}\right) \, \delta\bigl(\varepsilon - \varepsilon_0(1-k)^p(k/2)^l\bigl),
\end{align}
where in the course of derivation we made the substitution $s = i\omega$ and used a conventional integral representation for the Dirac $\delta$-function:
\begin{equation}
    \delta(x) = \cfrac{1}{2\pi}\int_{-\infty}^{+\infty}e^{i\omega x}d\omega.
\end{equation}
Similarly, for photon energy distribution we obtain 
\begin{equation}\label{f_g general}
    f_{\gamma}(\varepsilon, t) = \sum_{p, l = 0}^{\infty} B_{pl}\left(t, L_e, L_{\gamma}\right) \, \delta\bigl(\varepsilon - \varepsilon_0(1-k)^p(k/2)^lk\bigr).
\end{equation}

From Eqs.~\eqref{f_e general} and \eqref{f_g general} we can see that the cascade consists of leptons and photons with energies
\begin{eqnarray}
    \varepsilon_{pl}^{(e)} = \varepsilon_0(1-k)^p(k/2)^l, \label{e_spectrum} \\
    \varepsilon_{pl}^{(\gamma)} = \varepsilon_0(1-k)^p(k/2)^lk, \label{gamma_spectrum}
\end{eqnarray}
respectively. The coefficients $A_{pl}(t, L_e, L_{\gamma})$ and $B_{pl}(t, L_e, L_{\gamma})$ then represent the amounts of leptons and photons with the corresponding energies at depth $t$.
\begin{figure}[H]
    \centering
    \includegraphics[scale = 0.5]{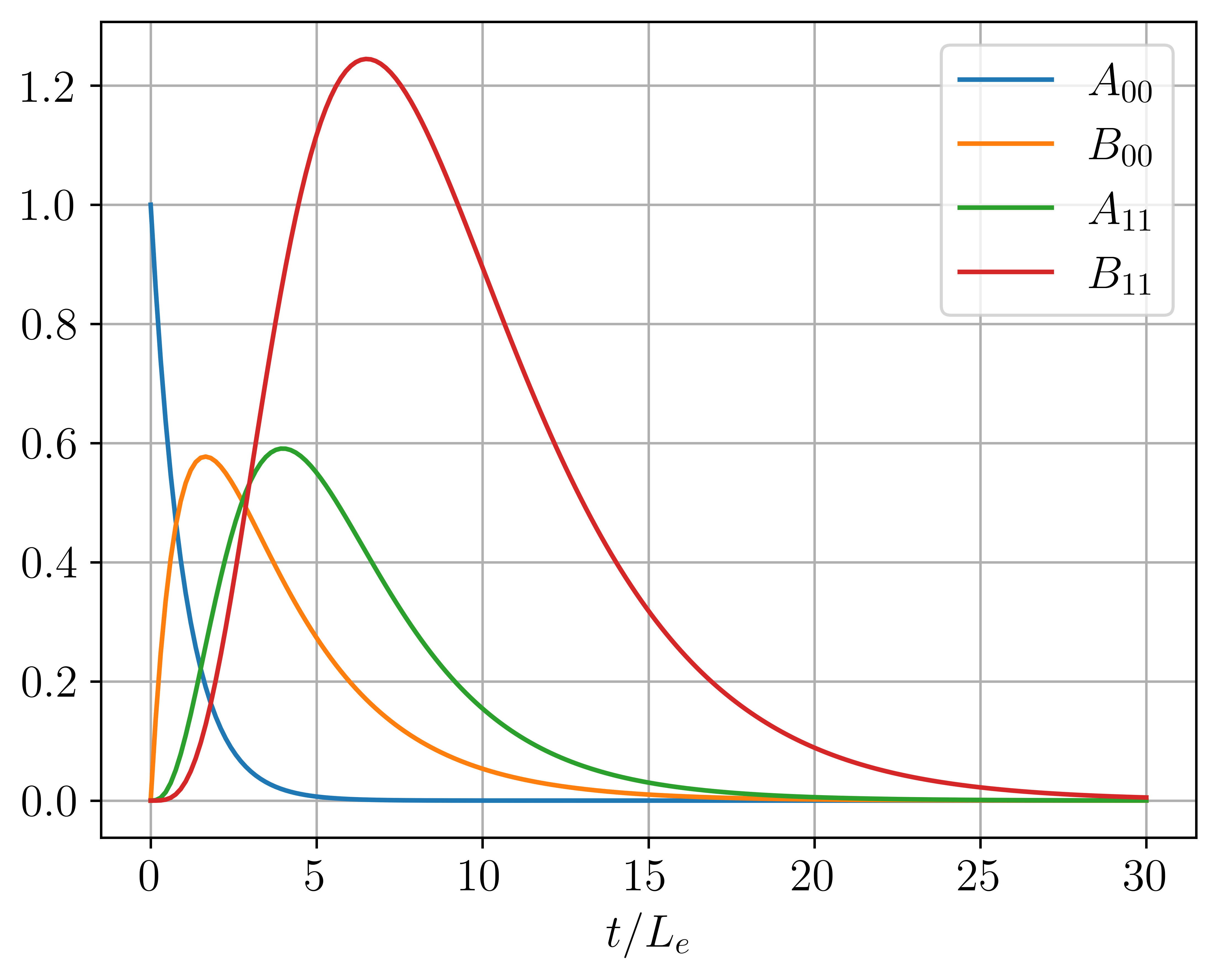}
    \caption{\label{fig:distr_func} The dependence of the coefficients $A_{00}, B_{00}, A_{11}$ and $B_{11}$ on depth $t$ for $L_e /L_{\gamma} = 1/3$.}
\end{figure}

As it is seen from Fig.~\ref{fig:distr_func}, for each energy the number of particles of that energy reaches maximum at certain depth and with further increase of depth tends to zero. For lower energies the maxima are stronger and are located at higher depth.

\section{Final number of leptons}

Our goal is to compute the final number of leptons produced in a QED cascade. Recall that Eqs.~\eqref{kinetic_equations} are literally valid only for $\varepsilon\geqslant E_0$. Therefore a lepton contributes to the final number of leptons when its energy falls below $E_0$. To track such leptons we count the number of processes with energies of initial particles greater than $E_0$ and of final leptons lower than $E_0$. Each hard photon emission results in one lepton and each photoproduction results in two leptons. Thus a general formula for the final number of leptons reads
\begin{eqnarray} \label{N_fin_general_formula}
        N_e = \int_0^{\infty} dt \int_{E_0}^{\infty} d\varepsilon \int_0^{E_0} \left[ f_e(\varepsilon, t) \, W_{e \rightarrow \gamma}(\varepsilon, \varepsilon - \varepsilon') +~2 \, f_{\gamma}(\varepsilon, t) \, W_{\gamma \rightarrow e}(\varepsilon, \varepsilon') \right] d\varepsilon'.
\end{eqnarray}
Substituting Eqs.~\eqref{W_eg} into Eq.~\eqref{N_fin_general_formula} we obtain
    \begin{eqnarray}\label{N_fin_computing_formula}
        N_e = \int_0^{\infty} \left[ \cfrac{1}{L_e} \int_{E_0}^{E_0 / (1-k)} f_e(\varepsilon, t) d\varepsilon + \cfrac{2}{L_{\gamma}} \int_{E_0}^{2E_0} f_{\gamma}(\varepsilon, t) d\varepsilon \right] dt.  
\end{eqnarray}

\par Note that Eq.~\eqref{N_fin_computing_formula} contains integration over depth and that distributions can be rewritten as inverse Mellin transform of the functions $\hat{f}_e(s, t)$ and $\hat{f}_{\gamma}(s, t)$ via Eq.~\eqref{Mellin transform inv}. Integrating them over depth $t$ we arrive at
\begin{equation}
\label{f_hat_integral_over_time}
\begin{aligned}
    \cfrac{1}{L_e} \int_0^{\infty} \hat{f_e}(s, t) dt= & ~ \cfrac{\varepsilon_0^s}{1 - (1-k)^s - 2 \, (k/2)^s}, \\
    \cfrac{1}{L_{\gamma}} \int_0^{\infty} \hat{f_{\gamma}}(s, t) dt= & ~ \cfrac{\varepsilon_0^s \, k^s}{1 - (1-k)^s - 2 \, (k/2)^s}.
\end{aligned}
\end{equation}
It is worth emphasizing that in Eqs.~\eqref{f_hat_integral_over_time} the free paths $L_e$ and $L_{\gamma}$ cancel out. This indicates that actually the final number of leptons should not depend on them. Indeed, this is explicitly shown  below in derivation of Eq.~\eqref{N_fin_integral} in Sec.~\ref{sec:appr}. Here, without loss of generality, let us just suppose for simplicity that $L_e = L_{\gamma} \equiv L$. Then Eq.~ \eqref{N_fin_computing_formula} takes the form
\begin{align}\label{N_fin_Le=Lg}
    \nonumber N_e = & \int_0^{\infty} \left[ \cfrac{1}{L} \int_{E_0}^{E_0 / (1-k)} f_e(\varepsilon, t) d\varepsilon + \cfrac{2}{L} \int_{E_0}^{2E_0} f_{\gamma}(\varepsilon, t) d\varepsilon \right] dt = \\ 
    = & \int_{E_0}^{E_0 / (1-k)} \left(\int_0^{\infty} \cfrac{f_e(\varepsilon, t)}{L} dt\right) d\varepsilon + 2 \, \int_{E_0}^{2E_0} \left(\int_0^{\infty} \cfrac{f_{\gamma}(\varepsilon, t)}{L} dt\right) d\varepsilon,
\end{align}
and the particle energy distributions given by Eqs.~\eqref{f_e general}, \eqref{Apl} and Eqs.~\eqref{f_g general}, \eqref{Bpl}, respectively, turn into
\begin{eqnarray}
    \label{f_e_L} f_e(\varepsilon, t) = e^{-t/L} \, \sum_{p, l = 0}^{\infty}\sum_{n=l}^{l + \lfloor p/2 \rfloor} \left( \cfrac{t}{2L} \right)^{p+2l} \cfrac{(p+2l+1) \, C_n^l \, 8^l}{(p - 2(n-l))!(2n+1)!} \, \delta\bigl(\varepsilon - \varepsilon_0(1-k)^p(k/2)^l\bigr), \\
    \label{f_g_L} f_{\gamma}(\varepsilon, t) = 2e^{-t/L} \sum_{p, l = 0}^{\infty} \sum_{n=l}^{l + \lfloor p/2 \rfloor}\left(\cfrac{t}{2L}\right)^{p+2l+1}\cfrac{C_n^l \, 8^l}{(p-2(n-l))!(2n+1)!} \, \delta\bigl(\varepsilon - \varepsilon_0(1-k)^p(k/2)^lk \bigr),
\end{eqnarray}
where $C_n^l = n! / l!(n-l)!$ are the binomial coefficients and $\lfloor x \rfloor$ denotes the floor function.

By substituting Eqs.~\eqref{f_e_L} and \eqref{f_g_L} into Eq.~\eqref{N_fin_Le=Lg} and
using the definition of the Euler $\Gamma$-function we can integrate over depth:
\begin{align}
    \label{eq36} \int_0^{\infty} \cfrac{f_e(\varepsilon, t)}{L} dt = & \sum_{p, l = 0}^{\infty} 2^{l-p} \, \left(\sum_{n=l}^{l + \lfloor p/2 \rfloor}  C_n^l \, C_{p+2l+1}^{2n+1} \right) \, \delta \bigl(\varepsilon - \varepsilon_0(1-k)^p(k/2)^l\bigr), \\
    \label{eq37} \int_0^{\infty} \cfrac{f_{\gamma}(\varepsilon, t)}{L} dt = & \sum_{p, l = 0}^{\infty} 2^{l-p} \, \left(\sum_{n=l}^{l + \lfloor p/2 \rfloor}  C_n^l \, C_{p+2l+1}^{2n+1} \right) \, \delta \bigl(\varepsilon - \varepsilon_0(1-k)^p(k/2)^lk\bigr).
\end{align}
Expressions \eqref{eq36} and \eqref{eq37} can be further simplified by using the combinatorial identity from \cite{H.W.Gould_Identities}:
\begin{equation}
    \sum^{\lfloor m / 2 \rfloor}_{n = l} C_n^l C_{m+1}^{2n+1} = 2^{m - 2l} C_{m-l}^l.
\end{equation}
Finally, combining all together we arrive at
\begin{equation}\label{Nfin_final_formula}
    N_e = \left(\sum_{E_0 \leqslant \varepsilon_0 (1-k)^p(k/2)^l \leqslant E_0 / (1-k)} + 2 \, \sum_{E_0 \leqslant \varepsilon_0 (1-k)^p(k/2)^lk \leqslant 2E_0} \right) 2^l C_{p+l}^l.
\end{equation}

\begin{figure}[t]
    \centering
    \subfloat{
    \includegraphics[scale = 0.5]{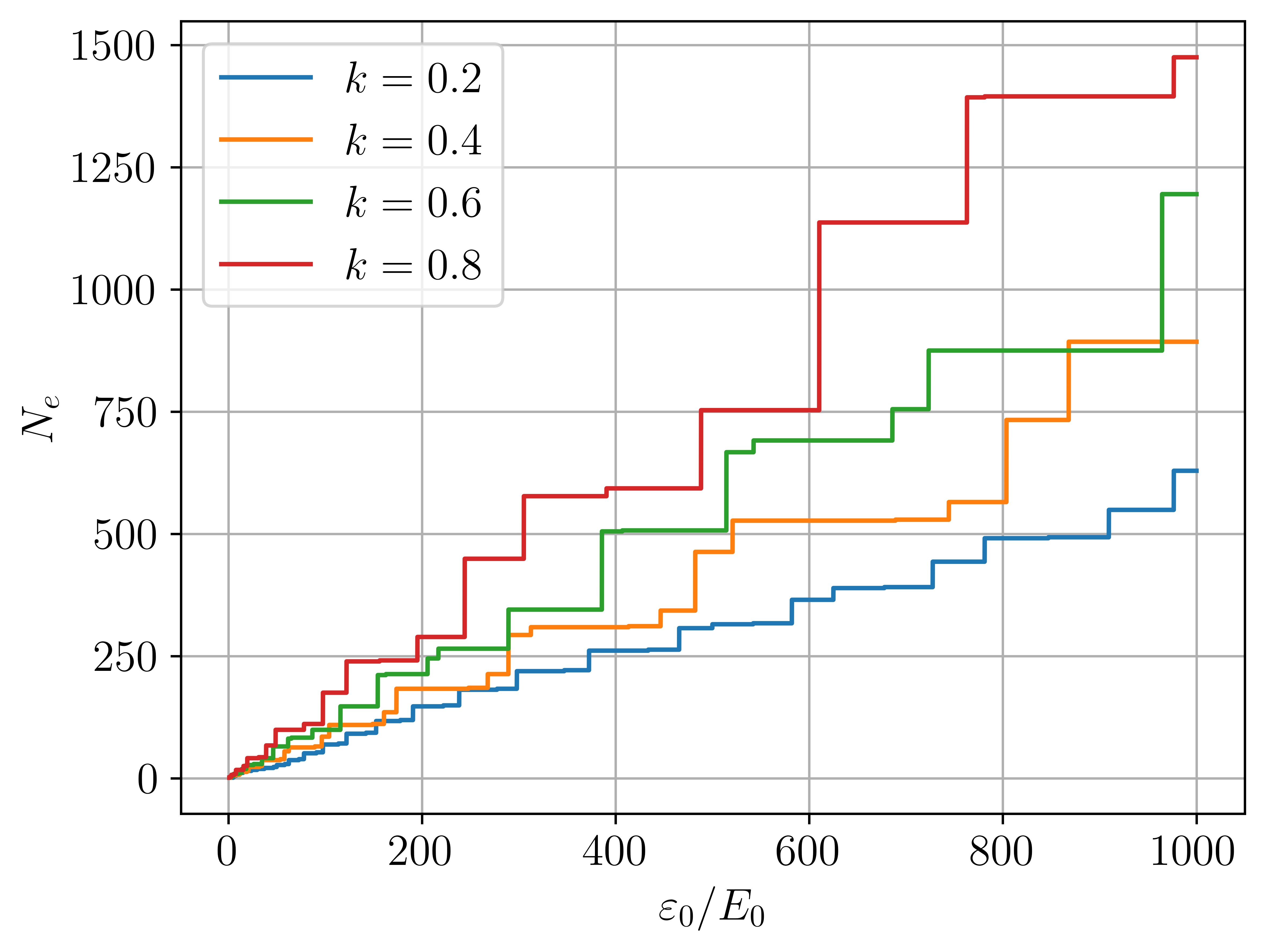}}\qquad\subfloat{
    \includegraphics[scale = 0.5]{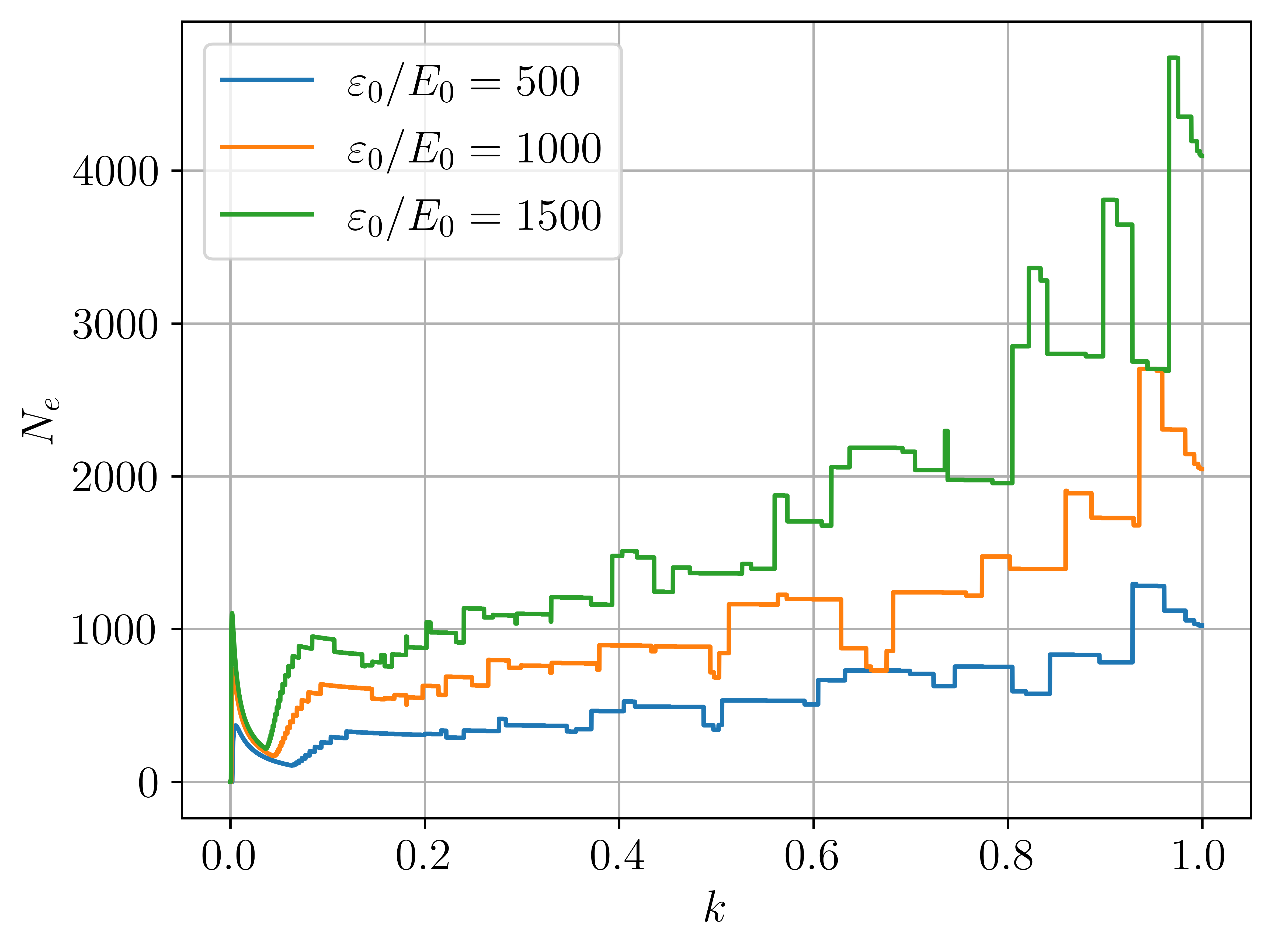}}
        \caption{\label{fig:N_to_epsilon0E0_k_exact} The final number of leptons as a function of $\varepsilon_0/E_0$ (left) and $k$ (right).}
\end{figure}

We emphasize that the resulting expression \eqref{Nfin_final_formula} for the final number of leptons is actually valid for any values of the free paths $L_e$ and $L_{\gamma}$. It appears to be a function $N_e = N_e(\varepsilon_0/E_0, k)$ of two dimensionless parameters, the ratio of the seed electron energy $\varepsilon_0$ to the threshold energy $E_0$ and the energy transfer coefficient $k$ for the photon emission process. The dependence of the final number of leptons on them is illustrated in Fig.~\ref{fig:N_to_epsilon0E0_k_exact}. In particular, it  has a stair-like growth with a constant average slope with increase of $\varepsilon_0$.

\section{Combinatorial interpretation}

Let us explain the obtained result \eqref{Nfin_final_formula} from combinatorial viewpoint. 
As discussed in Sec.~\ref{sec4}, in our model particles in the cascade have certain discrete energy values. The physical meaning behind these energy values is that $\varepsilon_{pl}^{(e)} = \varepsilon_0(1-k)^p(k/2)^l$ is the energy of leptons produced as a result of $l$ successive photon emissions followed by pair productions and $p$ additional photon emissions. The energy $\varepsilon_{pl}^{(\gamma)} = \varepsilon_0(1-k)^p(k/2)^lk$ is interpreted as the energy of a last photon emitted by a lepton with energy $\varepsilon_{pl}^{(e)}$. For each energy $\varepsilon_{pl}^{(e)}$ the total number of leptons of that energy can be found as a number of corresponding combinations of simple photon emissions and those with successive pair photoproduction with account for that each latter process doubles the number of leptons in the same combination. Obviously, the number of photons with energy $\varepsilon_{pl}^{(\gamma)}$ is the same:
\begin{equation}\label{Ne_Ng_number} 
    N_e(\varepsilon = \varepsilon_{pl}^{(e)}) = 2^lC_{p+l}^l, \quad N_{\gamma}(\varepsilon = \varepsilon_{pl}^{(\gamma)})= 2^lC_{p+l}^l.
\end{equation}

Each lepton with energy $E_0 \leqslant \varepsilon \leqslant E_0/(1-k)$ by emitting the last photon acquires energy $\varepsilon' \leqslant E_0$ thus adding one to the final number of leptons. Similarly, each photon with energy $E_0 \leqslant \varepsilon \leqslant 2 E_0$ produces a pair with energies below $E_0$. Therefore for the final number of leptons we have
\begin{align}\label{combinatorial_Nfin}
    N_e = & N_e(E_0 \leqslant \varepsilon \leqslant E_0/(1-k)) + 2 \, N_{\gamma}(E_0 \leqslant \varepsilon \leqslant 2E_0).
\end{align}
Taking into account Eqs.~\eqref{e_spectrum}, \eqref{gamma_spectrum} and expressions \eqref{Ne_Ng_number}, from Eq.~\eqref{combinatorial_Nfin} we arrive at Eq.~\eqref{Nfin_final_formula}.

\section{Approximate formula at high seed electron energies}
\label{sec:appr}

It is interesting to analyze the final number of leptons at high seed electron energies to identify the average slope of its dependence on $\varepsilon_0/E_0$. 
Let us start by obtaining an integral representation for the final number of leptons. Since the final number of leptons is a function of dimensionless parameters, it is convenient to introduce the dimensionless variables and distributions:
\begin{eqnarray}
\nonumber\varepsilon / \varepsilon_0 \equiv \xi, ~ E_0 / \varepsilon_0 \equiv \eta, \\ \label{dimensionless_notation}
    f(\varepsilon, t) \equiv \cfrac{1}{\varepsilon_0} \, g(\xi, t) \rightarrow \hat{f}(s, t) = \varepsilon_0^s \, \hat{g}(s, t).
\end{eqnarray}
With this notation Eq.~\eqref{N_fin_computing_formula} takes the form
\begin{equation}
    N_e = N_e(\eta, k) = \int_0^{\infty} \left[ \cfrac{1}{L_e} \int_{\eta}^{\eta / (1-k)} g_e(\xi, t) d\xi + \cfrac{2}{L_{\gamma}} \int_{\eta}^{2\eta} g_{\gamma}(\xi, t) d\xi \right] dt.
\end{equation}
The Mellin transform of $N_e(\eta, k)$ over $\eta$ according to  definition \eqref{Mellin transform} reads
\begin{align}
    \nonumber\hat{N}_e(s, k) &= \mathcal{M}[N_e](s) = \int_0^{\infty} \eta^s N_e(\eta, k) d\eta = \\
    &= \int_0^{\infty} \left[ \cfrac{1}{L_e} \int_0^{\infty} d\eta \, \eta^s \int_{\eta}^{\eta / (1-k)} g_e(\xi, t) d\xi + \cfrac{2}{L_{\gamma}} \int_0^{\infty} d\eta \, \eta^s \int_{\eta}^{2\eta} g_{\gamma}(\xi, t) d\xi \right] dt.
    \label{eq:n_e_dimless}
\end{align}
Let us transform the first term in the integral
\begin{align}
    \nonumber&\int_0^{\infty}\cfrac{dt}{L_e}\int_0^{\infty}d\eta \, \eta^s\int_{\eta}^{\eta / (1-k)} d\xi g_e(\xi, t) = \int_0^{\infty}\cfrac{dt}{L_e} \int_{1}^{1 / (1-k)} d\varkappa \int_0^{\infty}d\eta \, \eta^{s+1}g_e(\varkappa \eta, t) = \\
    &= \int_0^{\infty}\cfrac{dt}{L_e} \int_{1}^{1 / (1-k)} \cfrac{d\varkappa}{\varkappa^{s+2}}\int_0^{\infty}d\xi \, \xi^{s+1}g_e(\xi, t) = \cfrac{1 - (1-k)^{s+1}}{s+1}\int_0^{\infty}\cfrac{dt}{L_e} \, \hat{g}_e(s+1, t),
    \label{eq:1st_term}
\end{align}
where the change of integration variable $\xi = \varkappa \eta$ is made back and forth.
Using Eqs.~\eqref{dimensionless_notation} and \eqref{f_hat_integral_over_time} for Eq.~\eqref{eq:1st_term} we obtain
\begin{equation}
    \cfrac{1 - (1-k)^{s+1}}{s+1} \, \cfrac{1}{1 - (1-k)^{s+1} - 2 \, (k/2)^{s+1}}.
\end{equation}
After similar calculations with the second term for Eq.~\eqref{eq:n_e_dimless} we get
\begin{equation}
    \hat{N}_e(s, k) = \cfrac{1}{s+1} \, \cfrac{1 +2k^{s+1} - (1 - k)^{s+1} - 2(k / 2)^{s+1}}{1 - (1 - k)^{s+1} - 2(k / 2)^{s+1}}.
\end{equation}
The final number of leptons can be now recovered as an inverse Mellin transform of the obtained expression as follows:
\begin{equation}\label{N_fin_integral}
    N_e(\eta, k) = \cfrac{1}{2 \pi i}\int_{\sigma - i\infty}^{\sigma+i\infty}\cfrac{ds~ \eta^{-s-1}}{s+1} \, \cfrac{1 +2k^{s+1} - (1 - k)^{s+1} - 2(k / 2)^{s+1}}{1 - (1 - k)^{s+1} - 2(k / 2)^{s+1}}.
\end{equation}
\begin{figure}[t]
    \centering
    \includegraphics[scale = 0.5]{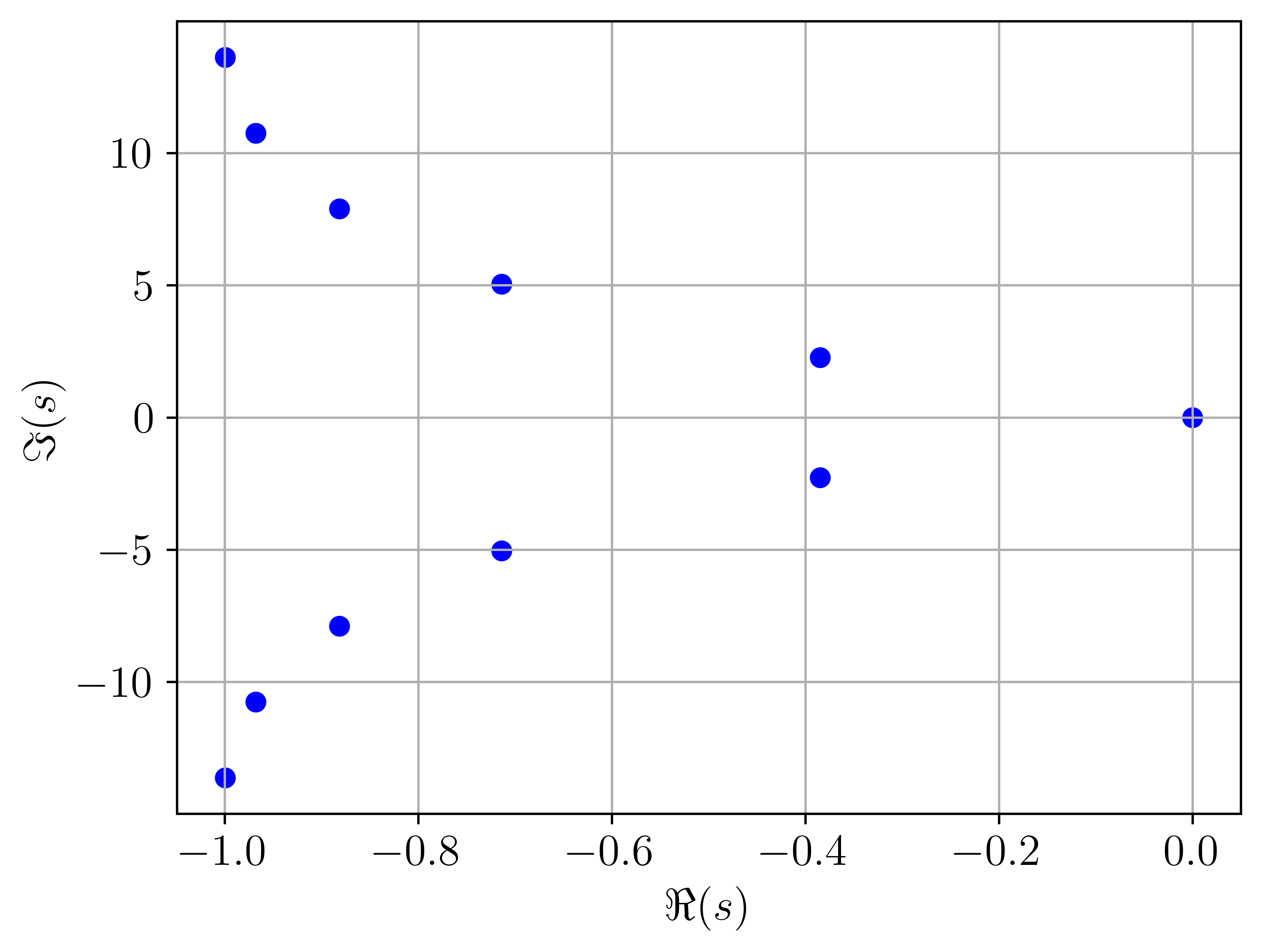}
    \caption{\label{fig:poles} The location of simple poles of $\hat{N}_e(s, k)$ in the $s$-plane at $k = 0.2$.}
\end{figure}
Function $\hat{N}_e(s, k)$ has a removable singularity at $s = -1$ and a set of simple poles defined by the equation
\begin{equation} \label{pole_eq}
    1 - (1-k)^{s+1} - 2(k/2)^{s+1} = 0.
\end{equation}
Representing complex variable $s = x + iy, ~ x, y \in \mathbb{R}$, Eq.~\eqref{pole_eq} is split in real and imaginary parts as
\begin{align}
    \label{real_part_eq}& (1-k)^{x+1} \, \cos(y \ln(1-k)) + 2(k/2)^{x+1} \, \cos(y \ln(k/2)) = 1, \\
    \label{imag_part_eq}& (1-k)^{x+1} \, \sin(y \ln(1-k)) + 2(k/2)^{x+1} \, \sin(y \ln(k/2)) = 0.
\end{align}
Let us estimate the left-hand side of Eq.~\eqref{real_part_eq}:
\begin{equation}\label{constr_long}
    (1-k)^{x+1} \, \cos(y \ln(1-k)) + 2(k/2)^{x+1} \, \cos(y \ln(k/2)) \leqslant (1-k)^{x+1} + 2(k/2)^{x+1}.
\end{equation}
For a solution of equation \eqref{real_part_eq} to exist it is necessary that
\begin{equation}\label{constraint}
    (1-k)^{x+1} + 2(k/2)^{x+1} \geqslant 1.
\end{equation}
Note that $s=0$ is a solution of Eq.~\eqref{pole_eq}. Since the left-hand side of inequality \eqref{constraint} is a monotonically decreasing function of $x$, from Eq.~\eqref{constraint} we get a constraint on the real parts of the simple poles:
\begin{equation} \label{x <= 0}
    x = \Re(s) \leqslant 0.
\end{equation}
Their location is illustrated in Fig.~\ref{fig:poles}.

Therefore we can evaluate the integral in Eq.~\eqref{N_fin_integral} by choosing any $\sigma > 0$ and using the residue theorem:
\begin{equation}\label{eq:n_fin_res}
        N_e = \sum_{\nu = -\infty}^{+\infty} \cfrac{1}{s_{\nu}+1} \left(\cfrac{\varepsilon_0}{E_0}\right)^{{s_{\nu}+1}} \cfrac{1 + 2 k^{{s_{\nu}+1}} - (1 - k)^{{s_{\nu}+1}} - 2(k / 2)^{{s_{\nu}+1}}}{(1 - k)^{{s_{\nu}+1}} \ln(1/(1-k)) + 2(k / 2)^{{s_{\nu}+1}} \ln(2/k)},
\end{equation}
where $s_{\nu}$ are all the poles numerated from bottom to top so that $s_0=0$. Since $s=0$ is among them, the final number of leptons always contains at least one term linear in $\varepsilon_0/E_0$. According to Eq.~\eqref{x <= 0}, at high seed electron energies $\varepsilon_0 \gg E_0$, such terms are the leading ones, so that Eq.~\eqref{eq:n_fin_res} can be written as
\begin{equation}
    N_e = N_e^{(1)} +  o(\varepsilon_0 / E_0),
\end{equation}
where $N_e^{(1)}$ collects the contribution of all poles with $\Re(s) = 0$ that are linear in $\varepsilon_0/E_0$. As already mentioned, at high energy they are leading in the expansion, hence can be used as a reasonable approximation.

At $x=0$ it follows from Eq.~\eqref{constr_long} that Eq.~ \eqref{real_part_eq} is satisfied only  when its left-hand side attains maximum, i.e. when $y$ satisfies the conditions:
\begin{equation}\label{eq_im_part}
    \begin{cases}
        y \ln(1-k) = 2 \pi m,~m \in \mathbb{Z}, \\
        y \ln(k / 2) = 2 \pi n, ~n \in \mathbb{Z}.
    \end{cases}
\end{equation}
It is easy to see that if $y$ satisfies Eq.~\eqref{eq_im_part} then Eq.~\eqref{imag_part_eq} also becomes a true identity. 

\begin{figure}[t]
    \centering
    \subfloat{
    \includegraphics[scale = 0.5]{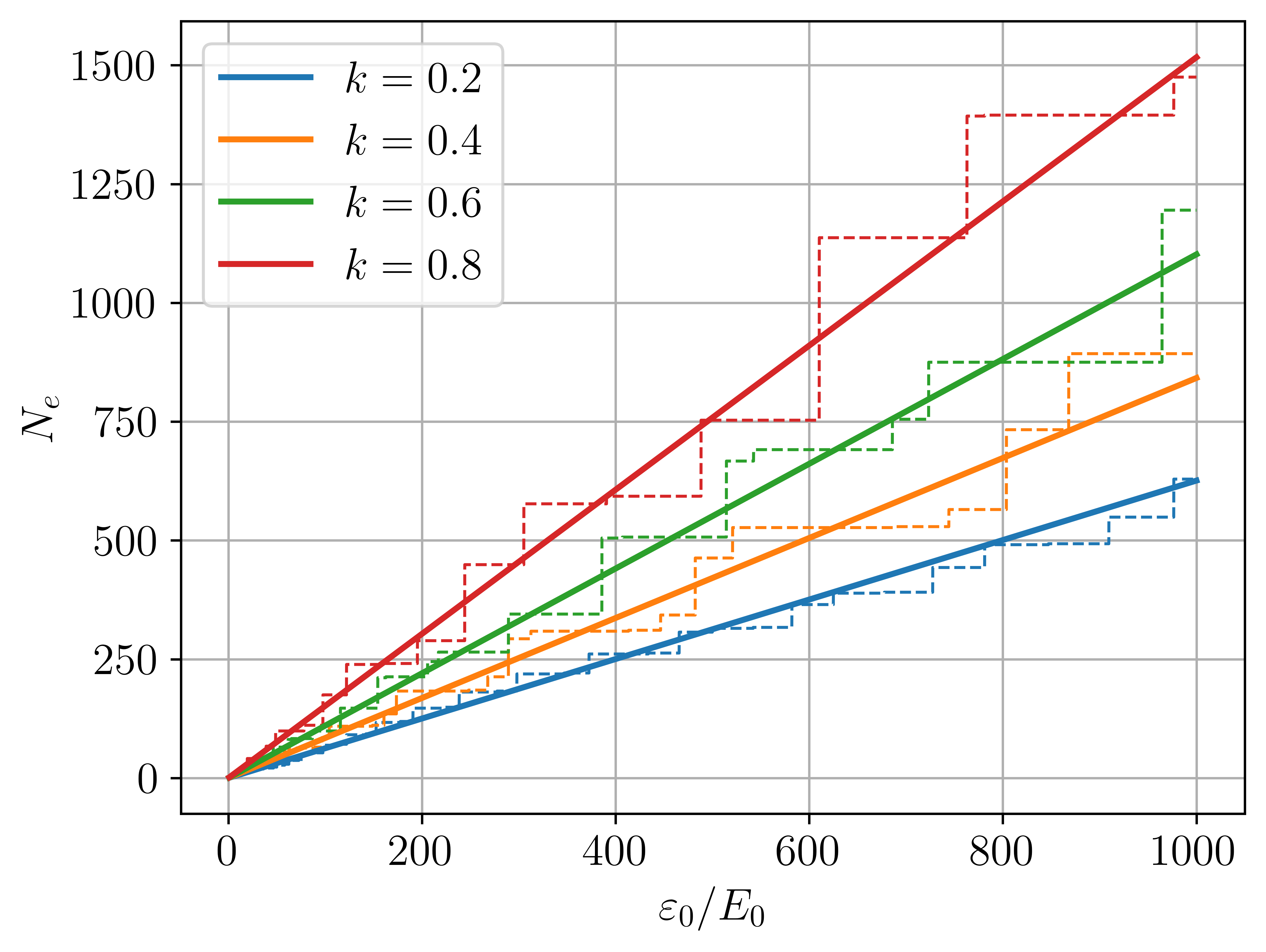}}\qquad
    \subfloat{
    \includegraphics[scale = 0.5]{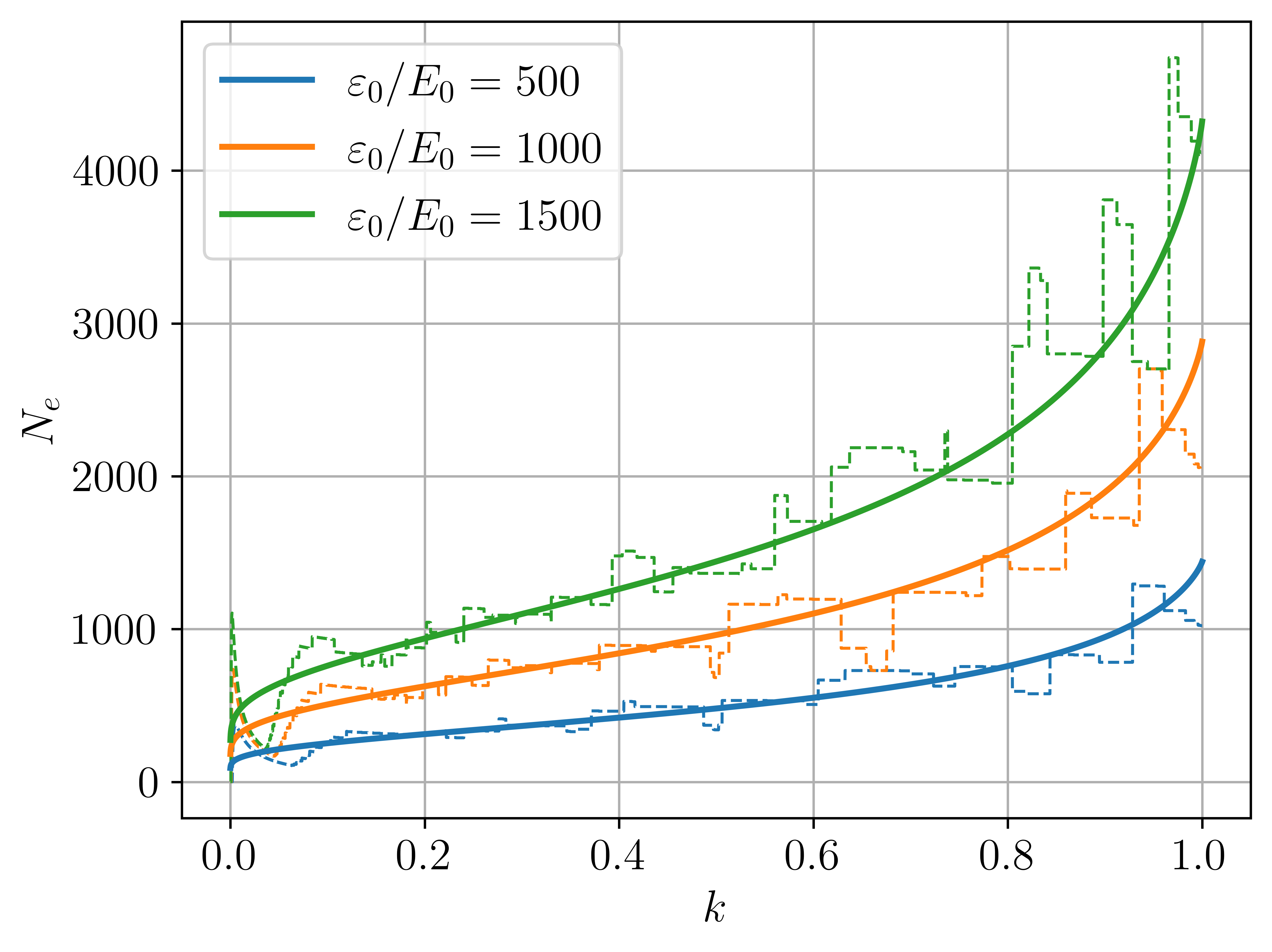}}
    \caption{\label{fig:N_to_epsilon0E0_k_exact_and_approx} The final number of leptons (dashed lines) and its linear approximation (solid lines) as functions of $\varepsilon_0/E_0$ (left) and $k$ (right).}
\end{figure}

Let us first consider the case of irrational $\ln(1-k) / \ln(k/2)$. Then $y = 0$ is the only solution of Eq.~ \eqref{eq_im_part} and $s=0$ is the only simple pole with $\Re(s) = 0$. The linear approximation for this case is
\begin{equation} \label{linear_irrational}
    N_e^{(1)} = \cfrac{\varepsilon_0}{E_0} \, \cfrac{2k}{(1-k)\ln(1 / (1-k)) + k\ln(2 / k)}.
\end{equation}
Exact solution \eqref{Nfin_final_formula} is compared to linear approximation \eqref{linear_irrational} in Fig.~ \ref{fig:N_to_epsilon0E0_k_exact_and_approx}.

For rational $\ln(1-k) / \ln(k/2)$, denote
\begin{equation}\label{ratio_log}
    \cfrac{\ln(1-k)}{\ln(k/2)} = \cfrac{m_0}{n_0},
\end{equation}
where $m_0/n_0$ is an irreducible fraction with $m_0, n_0 \in \mathbb{N}$. Then in Eq.~\eqref{eq_im_part} we have in general  $m = -\nu \, m_0,~ n = -\nu \, n_0, ~\nu \in \mathbb{Z}$. Note that $s_{\nu}=i\nu y_0$,
    \begin{eqnarray}
        (1-k)^{1+i\nu y_0} = (1-k) \, \exp\left(-i\nu\cfrac{2\pi m_0}{\ln(1-k)}\ln(1-k)\right) = (1-k), \\
        (k/2)^{1+i \nu y_0} = (k/2) \, \exp\left(-i\nu\cfrac{2\pi n_o}{\ln(k/2)}\ln(k/2)\right) = k/2, \\
        k^{1+i \nu y_0} = k \, \exp\left(-i\nu 2\pi n_o \cfrac{\ln(k) - \ln(2) + \ln(2)}{\ln(k) - \ln(2)}\right) = k \, e^{i \nu y_0 \ln(2)},
    \end{eqnarray}
where $y_0 = -2\pi m_0 / \ln(1-k) = -2\pi n_0 / \ln(k/2)$. 
    Then for the linear approximation we get:
    \begin{align} \label{linear_ratioanal_1}
         N_e^{(1)} & = \cfrac{\varepsilon_0 / E_0}{(1-k)\ln(1 / (1-k)) + k\ln(2 / k)} \sum_{\nu = -\infty}^{+\infty} \cfrac{2 k e^{i\nu y_o \ln(2\varepsilon_0 / E_0)}}{1 + i\nu y_0}.
    \end{align}
    In order to sum up the series we rewrite it in the following form:
    \begin{eqnarray}
        \nonumber\sum_{\nu = -\infty}^{+\infty} \cfrac{2 k e^{i\nu y_o \ln(2\varepsilon_0 / E_0)}}{1 + i\nu y_0} = 1 + 2 k \sum_{\nu = 1}^{+\infty}\left[ \cfrac{e^{i\nu y_o \ln(2\varepsilon_0 / E_0)}}{1 + i\nu y_0} + \cfrac{e^{-i\nu y_o \ln(2\varepsilon_0 / E_0)}}{1 - i\nu y_0} \right] = \\
        = 1 + \cfrac{2}{y_0^2} \, \sum_{\nu = 1}^{+\infty} \cfrac{\cos(\nu y_0 \ln(2\varepsilon_0/E_0))}{\nu^2 + 1/y_0^2} + \cfrac{2}{y_0} \, \sum_{\nu = 1}^{+\infty} \cfrac{\nu \sin(\nu y_0 \ln(2\varepsilon_0/E_0))}{\nu^2 + 1/y_0^2}.
    \end{eqnarray}
Note that since the shift $y_0\ln(2\varepsilon_0/E_0) \rightarrow y_0\ln(2\varepsilon_0/E_0) + 2\pi$ doesn't change the sum, $y_0\ln(2\varepsilon_0/E_0)$ can be replaced with $\bmod_{2\pi}[y_0\ln(2\varepsilon_0/E_0)]$. Then we can apply the formulas from Ref.~\cite{zwillinger2007table}:
    \begin{eqnarray}
        \sum_{\nu = 1}^{+\infty} \cfrac{\nu\sin(\nu x)}{\nu^2 + \alpha^2} = \cfrac{\pi}{2}\cfrac{\sinh\alpha(\pi-x)}{\sinh\alpha\pi},~ 0 < x < 2\pi, \\
        \sum_{\nu = 1}^{+\infty} \cfrac{\cos(\nu x)}{\nu^2 + \alpha^2} = \cfrac{\pi}{2\alpha}\cfrac{\cosh\alpha(\pi-x)}{\sinh\alpha\pi} - \cfrac{1}{2\alpha^2},~ 0 \leqslant x \leqslant 2\pi.
    \end{eqnarray}
    Finally, for the linear approximation we obtain
    \begin{equation}\label{linear_rational}
        N_e^{(1)} = \cfrac{\varepsilon_0}{E_0}\,\cfrac{2k}{(1-k)\ln(1 / (1-k)) + k\ln(2 / k)} \, \cfrac{\pi e^{\left\{\pi - \bmod_{2\pi}[y_0\ln(2\varepsilon_0/E_0)]\right\}/y_0}}{y_0\sinh(\pi/y_0)}.
    \end{equation}

In particular, for $k = 1/2$ (original Heitler model), we have 
\begin{equation}
    \cfrac{\ln(1-k)}{\ln(k/2)} = \cfrac{\ln(2)}{\ln(4)} = \cfrac{1}{2},
\end{equation}
and Eq.~\eqref{linear_rational} reduces to
\begin{equation}\label{Heitler_Ne_linear}
    N_e^{(1)} = \cfrac{2}{3} \, 2^{\lceil \log_2(\varepsilon_0 / E_0) \rceil},
\end{equation}
where $\lceil x\rceil$ is the ceil function, in accordance with Eq.~\eqref{Heitler_Ne}.

\section{Cascade depth}
Having obtained the exact and approximate expressions for the final number of leptons, we can also estimate the depth $t_m$ at which this number is achieved. Integrating cascade equations \eqref{kinetic_equations} over the energy $\varepsilon$, we arrive at
\begin{equation}\label{eq:n_eqs}
\begin{aligned}
    \cfrac{dn_e(t)}{dt} &= \cfrac{2}{L_{\gamma}}\, n_{\gamma}(t), \\
    \cfrac{dn_{\gamma}(t)}{dt} &= \cfrac{1}{L_e}\, n_e(t) - \cfrac{1}{L_{\gamma}} \, n_{\gamma}(t), \\
    n_e(0) &= 1,~ n_{\gamma}(0) = 0,
\end{aligned}
\end{equation}
where $n_e(t)=N_e(+\infty,t)$ and $n_{\gamma}(t)=N_\gamma(+\infty,t)$ are the total numbers of leptons and photons, respectively. Even though Eqs.~\eqref{kinetic_equations} are valid only at $\varepsilon_0 \geqslant E_0$, here we integrated over the energies from zero to infinity. In doing so we ignored that in our model all the photons with energies lower than $E_0$ do not produce new pairs. The typical energies of the particles in the cascade go down to $E_0$ at $t\sim t_m$ . Therefore Eqs.~\eqref{eq:n_eqs} are approximately valid but only for $t\lesssim t_m$.

The solutions for $n_e(t)$ and $n_{\gamma}(t)$ read:
\begin{align}
    \label{n_e(t)}n_e(t) &= \cfrac{1}{2}\left( 1 + \cfrac{1}{\sqrt{1+8L_{\gamma}/L_e}} \right)e^{\mu_1 t} + \cfrac{1}{2}\left( 1 - \cfrac{1}{\sqrt{1+8L_{\gamma}/L_e}} \right)e^{\mu_2 t}, \\
    n_{\gamma}(t) &= \cfrac{L_{\gamma} / L_e}{\sqrt{1+8L_{\gamma}/L_e}}\left( e^{\mu_1 t} - e^{\mu_2 t} \right), \\
    \mu_{1,2} &= \cfrac{1}{2L_{\gamma}}\left( -1 \pm \sqrt{1 + 8L_{\gamma}/L_e} \right).
\end{align}

At high seed electron energy $\varepsilon_0$ the final number of leptons $N_e$ and the depth $t_m$ at which it is achieved are also high. By neglecting the decreasing exponential in Eq.~\eqref{n_e(t)} we arrive at an equation for $t_m$:
\begin{equation}
    n_e(t_m) \approx \cfrac{1}{2}\left( 1 + \cfrac{1}{\sqrt{1+8L_{\gamma}/L_e}} \right)e^{\mu_1 t_m} = N_e \approx N_e^{(1)}.
\end{equation}
Its solution reads
\begin{equation} 
\label{tm_linear}
t_m \approx \cfrac{2L_{\gamma}}{\sqrt{1+8L_{\gamma}/L_e}-1} \, \ln\left( \cfrac{2N_e^{(1)}}{1+1/\sqrt{1+8L_{\gamma}/L_e}} \right). 
\end{equation}
As discussed after Eqs.~\eqref{eq:n_eqs}, this solution lies at the bound of validity of Eqs.~\eqref{eq:n_eqs}, hence can give only a rough estimation. Nevertheless, in the original Heitler model with account for Eq.~\eqref{Heitler_Ne_linear} expression \eqref{tm_linear} reduces to
\begin{equation}
    t_m \approx L \ln(2) \, \left\lceil \cfrac{\ln(\varepsilon_0/E_0)}{\ln(2)} \right\rceil,
\end{equation}
in a perfect agreement with Eq.~\eqref{Heitler_tm}.

\section{Summary}

In this paper we have presented a generalized Heitler model for a QED cascade with the arbitrary fixed energy transfer coefficient for photon emission and different free paths of leptons and photons. 

We have found analytical solutions for the energy distributions above the photoproduction threshold, as well as the exact and approximate expressions for the final number of leptons in a cascade. We have proved that in this model the final number of leptons is independent on the free paths and asymptotically behaves as a linear function of the seed electron energy. Another useful characteristic, the cascade depth, has been roughly estimated. The results for the original Heitler model are reproduced as a special case.

Our findings can be used for the analysis of realistic Monte-Carlo simulations, in particular, of shower-type cascades developing in strong laser fields, which is at the moment a hot topic \cite{pouyez2024multiplicity,qu2024creating,tang2024finite}.

\acknowledgments

The work was supported by the MEPhI Program Priority 2030 and by the Foundation for the Advancement of Theoretical Physics and Mathematics ``BASIS'' (Grant No. 24-1-1-21-1).

\appendix
\section{Explicit forms of coefficients $\bm{A_{pl}, B_{pl}}$}
\label{sec:app}
Let us rewrite the expression for $\hat{f}_e(s, t)$:
\begin{align}\label{eq:a_init}
        \hat{f}_e(s, t) = \cfrac{\varepsilon_0^s}{R}\,  \exp\left[\cfrac{t}{2}\left(\cfrac{(1-k)^s}{L_e} - \cfrac{L_e+L_{\gamma}}{L_e L_{\gamma}}\right)\right] \, \left[\left(\cfrac{(1-k)^s-1}{L_e} + \cfrac{1}{L_{\gamma}}\right)  \sinh\left(\frac{Rt}{2}\right) +  R  \cosh\left(\frac{Rt}{2}\right) \right],
\end{align}
where for the sake of bravity we introduced a shortcut
$$
R=\sqrt{\left( \cfrac{(1-k)^s-1}{L_e} + \cfrac{1}{L_{\gamma}}\right)^2 + \cfrac{8}{L_e L_{\gamma}}\,\left(\cfrac{k}{2}\right)^s}.
$$
Let us expand Eq.~\eqref{eq:a_init} in powers of $t$:
    \begin{align}
        \nonumber \hat{f}_e(s, t) = & \varepsilon_0^s\exp\left(-\cfrac{L_e + L_{\gamma}}{2L_e L_{\gamma}}t\right) \, \left[ \sum_{m=0}^{\infty}\left(\cfrac{t}{2L_e}\right)^m \cfrac{(1-k)^{ms}}{m!} \right] \, \left[ \left(\cfrac{(1-k)^s-1}{L_e} + \cfrac{1}{L_{\gamma}}\right) \sum_{n=0}^{\infty} \left(\cfrac{t}{2}\right)^{2n+1} \cfrac{R^{2n}}{(2n+1)!} \right. \\
        & \left.+ \sum_{n=0}^{\infty} \left(\cfrac{t}{2}\right)^{2n} \cfrac{R^{2n}}{(2n)!}\right], \label{eqA5}\end{align}
where
\begin{align}
        \nonumber R^{2n} = & \left[ \left( \cfrac{(1-k)^s-1}{L_e} + \cfrac{1}{L_{\gamma}}\right)^2 + \cfrac{8}{L_e L_{\gamma}}\,\left(\cfrac{k}{2}\right)^s \right]^{n} = \sum_{l=0}^{n}C_n^l\left(\cfrac{8}{L_e L_{\gamma}}\right)^l \left(\cfrac{k}{2}\right)^{ls}\left( \cfrac{(1-k)^s-1}{L_e} + \cfrac{1}{L_{\gamma}}\right)^{2(n-l)} = \\
         = & \sum_{l=0}^{n}\sum_{j=0}^{2(n-l)}C_n^l C_{2(n-l)}^j \left(\cfrac{8}{L_e L_{\gamma}}\right)^l \left(\cfrac{L_e - L_{\gamma}}{L_{\gamma}}\right)^{2(n-l)-j}\cfrac{1}{L_e^{2(n-l)}}~(1-k)^{js}(k/2)^{ls}. \label{eqA6}
    \end{align}
Combining Eqs.~\eqref{eqA5} and \eqref{eqA6} and rearranging the terms we get:
    \begin{align}
        \nonumber \hat{f}_e(s, t) = & \varepsilon_0^s\exp\left(-\cfrac{L_e + L_{\gamma}}{2L_e L_{\gamma}}t\right)  \sum_{m=0}^{\infty}\sum_{n=0}^{\infty}\sum_{l=0}^{n}\sum_{j=0}^{2(n-l)} \left[ \left( \cfrac{(t/2)^{m+2n}}{m!(2n)!} \right. \right. \\
         &+\left.\left. \cfrac{L_e-L_{\gamma}}{L_e L_{\gamma}} \cfrac{(t/2)^{m+2n+1}}{m!(2n+1)!} \right) C_n^l C_{2(n-l)}^j \cfrac{(L_e-L_{\gamma})^{2(n-l)-j}8^l}{L_e^{m+2n-l}L_{\gamma}^{2n-l-j}}\, (1-k)^{m+j}(k/2)^l \right. \\
         & \left. + \cfrac{(t/2)^{m+2n+1}}{m!(2n+1)!} \, C_n^l C_{2(n-l)}^j \cfrac{(L_e-L_{\gamma})^{2(n-l)-j}8^l}{L_e^{m+2n-l+1}L_{\gamma}^{2n-l-j}}\, (1-k)^{m+j+1}(k/2)^l \right].
    \end{align}
Next we change the order of summation and introduce a new index $p = m + j$:
    \begin{eqnarray}
        \sum_{m=0}^{\infty}\sum_{n=0}^{\infty}\sum_{l=0}^{n}\sum_{j=0}^{2(n-l)} = \sum_{m=0}^{\infty}\sum_{l=0}^{\infty}\sum_{n=l}^{\infty}\sum_{j=0}^{2(n-l)} = \sum_{m=0}^{\infty}\sum_{l=0}^{\infty}\sum_{j=0}^{\infty}\sum_{n=l+\lceil j/2 \rceil}^{\infty} = \sum_{p=0}^{\infty}\sum_{l=0}^{\infty}\sum_{j=0}^{p}\sum_{n=l+\lceil j/2 \rceil}^{\infty}.
    \end{eqnarray}
Rearranging the terms we get:
    \begin{align}
        \nonumber \hat{f}_e(s, t) = & \varepsilon_0^s\exp\left(-\cfrac{L_e + L_{\gamma}}{2L_e L_{\gamma}}t\right) \, \sum_{p=0}^{\infty}\sum_{l=0}^{\infty}\sum_{j=0}^{p}\sum_{n=l+\lceil j/2 \rceil}^{\infty} \left[  
        \left(\cfrac{(t/2)^{p+2n-j}(p+2n-j+1)}{(2n+1)!(p-j)!} \right.\right. \\
        & \left.\left.+ \cfrac{L_e-L_{\gamma}}{L_e L_{\gamma}}\,\cfrac{(t/2)^{p+2n-j+1}}{(2n+1)!(p-j)!}\right) C_n^lC_{2(n-l)}^j \left( \cfrac{L_e-L_{\gamma}}{L_e L_{\gamma}} \right)^{2(n-l)-j} \cfrac{8^l}{L_e^{p+l}L_{\gamma}^l} \, (1-k)^{ps}(k/2)^{ls} \right].
    \end{align}
Changing the order of summation once again and introducing a new index $i = 2(n-l) - j$
    \begin{eqnarray}
        \sum_{j=0}^{p}\sum_{n=l+\lceil j/2 \rceil}^{\infty} = \sum_{n=l}^{\infty}\sum_{j=0}^{\min\{2(n-l), p\}} = \sum_{n=l}^{\infty}\sum_{i = \max\{2(n-l)-p, 0\}}^{2(n-l)},
    \end{eqnarray}
we obtain
    \begin{align}\label{eqA11}
        \nonumber \hat{f}_e(s, t) = & \varepsilon_0^s\exp\left(-\cfrac{L_e + L_{\gamma}}{2L_e L_{\gamma}}t\right) \, \sum_{p=0}^{\infty}\sum_{l=0}^{\infty}\sum_{n=l}^{\infty}\sum_{i = \max\{2(n-l)-p, 0\}}^{2(n-l)} \Bigg[  
        \bigg((t/2)^{p+2l+i}(p+2l+i+1) \\
        &  + \cfrac{L_e-L_{\gamma}}{L_e L_{\gamma}}\, (t/2)^{p+2l+i+1}\bigg) \cfrac{C_n^l C_{2(n-l)}^i8^l}{(2n+1)!(p-2(n-l)+i)!\, L_e^{p+l}L_{\gamma}^{l}} \left( \cfrac{L_e-L_{\gamma}}{L_e L_{\gamma}} \right)^{i} (1-k)^{ps}(k/2)^{ls} \Bigg].
    \end{align}
 Finally, comparing Eq.~\eqref{eqA11} to Eq.~\eqref{f_hat_e_expansion} we obtain the explicit form of the coefficients $A_{pl}(t, L_e, L_{\gamma})$:
    \begin{align}\label{Apl}
        \nonumber A_{pl}(t, L_e, L_{\gamma}) =& \exp\left(-\cfrac{L_e + L_{\gamma}}{2L_e L_{\gamma}}t\right) \, \sum_{n=l}^{\infty}\sum_{i = \max\{2(n-l)-p, 0\}}^{2(n-l)} \Bigg[  
        \bigg((t/2)^{p+2l+i}(p+2l+i+1) \\
        & + \cfrac{L_e-L_{\gamma}}{L_e L_{\gamma}}\, (t/2)^{p+2l+i+1}\bigg) \cfrac{C_n^l C_{2(n-l)}^i8^l}{(2n+1)!(p-2(n-l)+i)!\, L_e^{p+l}L_{\gamma}^{l}} \left( \cfrac{L_e-L_{\gamma}}{L_e L_{\gamma}} \right)^{i} \Bigg].
    \end{align}
The same way, for the coefficients $B_{pl}(t, L_e, L_{\gamma})$ we obtain
    \begin{align}\label{Bpl}
        B_{pl}(t, L_e, L_{\gamma}) = 2 \exp\left(-\cfrac{L_e + L_{\gamma}}{2L_e L_{\gamma}}t\right) \, \sum_{n=l}^{\infty}\sum_{i = \max\{2(n-l)-p, 0\}}^{2(n-l)}  \cfrac{(t/2)^{p+2l+i+1} \, C_n^l C_{2(n-l)}^i8^l}{(2n+1)!(p-2(n-l)+i)!\, L_e^{p+l+1}L_{\gamma}^{l}} \left( \cfrac{L_e-L_{\gamma}}{L_e L_{\gamma}} \right)^{i}.
    \end{align}

\providecommand{\noopsort}[1]{}\providecommand{\singleletter}[1]{#1}%
\end{document}